# Intermixing Induced Anisotropy Variations in CoB-based Chiral Multilayer Films

H. K. Tan[1], Royston J. J. Lim[1], H. L. Seng[1], J. Shanmugam[1], H. Y. Y. Ko[1], X. M. Cheng[3], V. Putra[1,2], Z. X. Xing[1], Anjan Soumyanarayanan[1,3†] and Pin Ho[1†]

[1] *Institute of Materials Research & Engineering, Agency for Science, Technology & Research (A*STAR), Singapore 138634*
[2] *Materials Science and Engineering Department, National University of Singapore, Singapore 117575*
[3] *Physics Department, National University of Singapore, Singapore 117551*

We examine the atomic intermixing phenomenon in three distinct amorphous CoB-based multilayer thin film platforms — Pt/CoB/Ir, Ir/CoB/Pt and Pt/CoB/MgO — which are shown to stabilise room-temperature chiral magnetic textures. Intermixing occurs predominantly between adjacent metallic layers. Notably, it is stack-order dependent, and particularly extensive when Ir sits atop CoB. Intermixing induced variations in magnetic properties are ascribed to the formation of magnetic dead layer arising from CoIr alloying in the metallic stacks. It also produces systematic variations in saturation magnetization, by as much as 30%, across stacks. Crucially, the resulting crossover CoB thickness for the transition from perpendicular to in-plane magnetic anisotropy differs by more than 2× across the stacks. Finally, with thermal annealing treatment over moderate temperatures of 150–300 °C, the magnetic anisotropy increases monotonically across all stacks, coupled with discernibly larger $H_c$ for the metallic stacks. These are attributed to thermally induced CoPt alloying and MgO crystallization in the metallic and oxide stacks, respectively. Remarkably, the CoB in the Pt/CoB/MgO stacks retains its amorphous nature after annealing. Our results set the stage for harnessing the collective attributes of amorphous CoB-based material platforms and associated annealing processes for modulating magnetic interactions, enabling the tuning of chiral magnetic texture properties in ambient conditions.

## 1. Introduction

Multilayer films containing heavy metal (HM)/ferromagnet (FM) interfaces are known to stabilize chiral magnetic textures such as domain walls and skyrmions at room-temperature (RT). Such chiral textures are of immense interest for memory, logic and brain-inspired computing applications.[1–5] Within chiral multilayers, spin–orbit coupling (SOC) under broken inversion symmetry produces the Dzyaloshinskii–Moriya interaction (DMI), which results in the canting of magnetic spins, with fixed handedness, known as chirality.[1,4,5] DMI competes with the conventional exchange interaction, or stiffness ($A$), which promotes parallel alignment of spins. Equally important is the effective magnetic anisotropy ($K_{\mathrm{eff}}$) which defines the energetically favourable spin orientation.[3,6,7] The $K_{\mathrm{eff}}$ plays a vital role in determining the chiral texture properties such as their stability, size, density, and dynamics. Controlling $K_{\mathrm{eff}}$ is crucial towards materials efforts for myriad of device applications.[6–9]

Conventionally, the $K_{\mathrm{eff}}$ of magnetic thin films include contributions from three components, namely magnetocrystalline anisotropy ($K_{\mathrm{u}}$), shape anisotropy ($2\pi M_{\mathrm{S}}^2$) and interfacial anisotropy ($K_{\mathrm{i}}$), given as[10,11]

$$K_{\mathrm{eff}} = K_u - 2\pi M_{\mathrm{s}}^2 + \frac{K_{\mathrm{i}}}{t} \qquad (1)$$

where shape anisotropy in a thin film is in-plane with a negative convention, $M_{\mathrm{s}}$ is the saturation magnetization and $t$ is the FM thickness. In HM/FM-based multilayer films, the out-of-plane $K_{\mathrm{eff}}$ arises predominantly from the interfacial magnetic anisotropy, which originates from the interfacial symmetry breaking-induced SOC, and the hybridization between the FM-3$d$ and HM-5$d$/O-2$p$ electron orbitals at the interfaces.[12–15] Finally, the shape anisotropy, determined by $M_{\mathrm{s}}$, is also a relevant factor in influencing $K_{\mathrm{eff}}$.

Earlier works have reported the tuning of stack anisotropy based on materials selection, such as the choice of interfacing HM or oxide,[16–18] multilayer stack ordering,[18] and layer thicknesses.[16,18,19] Additionally, deposition process relevant factors such as sputtering conditions and post-annealing result in intermixing and inter-diffusion induced alloying,[20,21] interfacial roughness[22] and crystallization effects,[19,23] which also influence the magnetic properties. In this light, we focus on amorphous CoB-based multilayer films as potential pin-free material platforms for energy-efficient addressability and manipulation of RT chiral textures.[24–26] While there have been earlier reports on the amorphous nature of CoB films,[27–29] studies on CoB-based multilayer platforms have been limited.[26] Crucially, the effects of adjacent HM or oxide layer, stack ordering, FM thickness and post-annealing on the magnetic properties of such CoB-based multilayer stacks — remain to be established.

In this work, we report on three distinct CoB-based multilayer material platforms: Pt/CoB/Ir, Ir/CoB/Pt and Pt/CoB/MgO, hosting RT skyrmion textures with sizes below 150 nm stabilized across the stacks. Interestingly, perpendicular magnetic anisotropy (PMA) is retained for the Pt/CoB/Ir multilayer up to an unexpectedly large CoB thickness of 2.8 nm, considerably greater than the crossover thicknesses of 1.9 and 1.3 nm for the Ir/CoB/Pt and Pt/CoB/MgO multilayer, respectively. We find that this difference arises from variances in the extent of intermixing across different CoB-based multilayer structures, and in stack ordering, leading to substantial variation of magnetic properties across such stacks. Further, post-annealing enables PMA enhancement across multilayers due to the formation of CoPt alloy in the metallic multilayer and crystallization of MgO in Pt/CoB/MgO. Our results correlate material stack design and annealing effects to the magnetic, crystallographic and microstructural properties of the CoB-based multilayers, providing a means of tuning $K_{eff}$ for engineering CoB-based multilayer well-suited to a range of device applications.

## 2. Methods

### *2.1 Film deposition and Characterization*

Multilayer stacks comprising a single CoB layer —

- Ta(3)/**Pt(5)/CoB(2–3.6)/Ir(1)**/Pt(4),
- Ta(3)/Pt(5)/**Ir(1)/CoB(1–2)/Pt(5),**
- Ta(3)/**Pt(5)/CoB(1–2)/MgO(1.5)**/Pt(4),

(nominal layer thicknesses in nm in parentheses), as well as corresponding stacks comprising of four repeats of the active layers (shown in bold), were deposited on pre-cleaned thermally oxidized 200 mm Si wafers by ultrahigh vacuum magnetron sputtering using the Timaris Singulus™ system. Additionally, a symmetric multilayer stack comprising Ta(3)/Pt(5)/CoB(1–2)/Pt(5) was deposited to serve as a reference against the three asymmetric chiral multilayer stacks (See S3). A composite target of $Co_{80}B_{20}$ was used for the CoB layer deposition (see S1). Deposition powers and pressures for Ta (2.5 kW, 200 sccm), Pt (0.075 W, 400 sccm), CoB (0.5 kW, 300 sccm), Ir (0.3 W, 300 sccm) and MgO (4 kW, 600 sccm) were identical across stacks. The single CoB layer stacks, with CoB thickness of 2 nm, were subsequently annealed at temperatures ranging from 150 to 300 °C for one hour in a magnetic vacuum annealing oven (Model PF184841) manufactured by Futek. Magnetization measurements were performed on the film using the MicroMag Model 2900™ alternating gradient magnetometer. The $M_s$, coercivity ($H_c$) and $K_{eff}$ were determined from the $M(H)$ data. The $K_{eff}$ of the films are derived from the areal difference between the out-of-plane (OP) and in-plane (IP) $M(H)$ loops.[11]

### *2.2 MFM Setup*

Magnetic force microscopy (MFM) imaging was performed using the Bruker Dimension ICON™ scanning probe microscope. To ensure high-resolution MFM images with minimal stray field perturbations, we adopt Co-alloy coated SSS-MFMR™ tips with sharp profile (diameter ~ 30 nm) and ultra-low moment (~ 80 emu $cm^{-3}$) at scanning lift heights of 20–30 nm. In our earlier works, we have established MFM as a reliable tool for imaging sub-100 nm skyrmions in multilayer films.[6,7] Here the multilayer stacks were imaged under *in situ* OP fields ranging from 0 to 80 mT, following *ex situ* negative OP field saturation.

### *2.3 TOF-SIMS*

Time-of-flight secondary ion mass spectrometry (TOF-SIMS) depth profiles of the multilayer stacks were acquired with both positive and negative polarities using TOF.SIMS 5 IONTOF GmbH instrument operating in dual beam mode. 30 keV pulsed $Bi^+$ primary ion beam was used for analysis over an area of 100 µm × 100 µm. Etching of the sample over a 500 µm × 500 µm area was performed using 500 eV $O_2^+$ or $Cs^+$ sputter ion beam, for positive or negative polarity, respectively. Here, only the depth profiles, obtained in the negative polarity mode, for characteristic Co, Pt, Ir and MgO secondary ions signals have been examined . Meanwhile, B profile was not shown due to its low detected counts in the negative polarity (see S3). The depth profiles were normalized point-to-point to the total ion intensity.

### *2.4 XRD*

X-ray diffraction (XRD) studies were performed on a Panalytical X'Pert PRO MRD X-ray diffractometer with $CuK_\alpha$ radiation (1.5406 Å). XRD scans were obtained over a scan range of 20–80° with step sizes of 0.05°. The size of the crystallites in the alloy was calculated from the broadening of the Bragg peak using the Scherrer equation, defined as[30]

$$d = \frac{k\lambda}{\beta cos\theta} \quad (2)$$

where $d$ is the crystallite size, $k$ is the shape factor or Scherrer constant assumed to be 1, $\lambda$ is the wavelength of the Cu source, $\beta$ and $\theta$ are the full width at half maximum (in radians) and position of the Bragg peak, respectively.

## 3. Chiral Multilayer Stack

Fig. 1(a)–(c) insets show the schematic of the three CoB-based multilayer material platforms for hosting RT chiral spin textures including stripes and skyrmions: Pt/CoB/Ir, Ir/CoB/Pt and Pt/CoB/MgO, resulting from the DMI at the respective FM (CoB) / HM (Pt, Ir) interfaces.[6,24,31] Fig. 1(a)–(c) show the out-of-plane (OP) magnetization hysteresis loop of representative samples of Pt/CoB(**2.4**)/Ir, Ir/CoB(**2.0**)/Pt and Pt/CoB(**1.0**)/MgO, with four stack repetitions used to ensure sufficient MFM signal. All three samples display near-zero

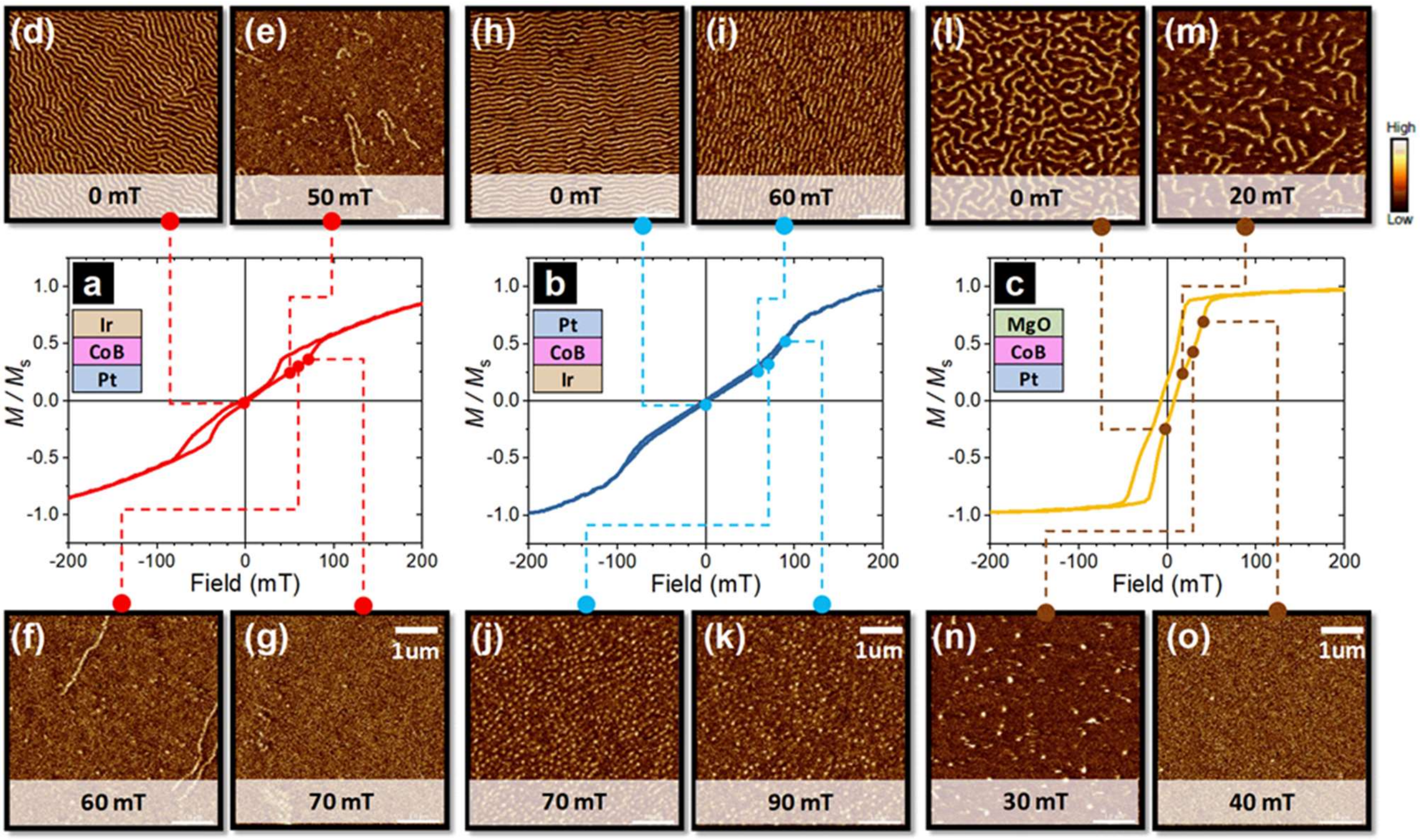

**Figure 1. MFM imaging of multilayer stacks. (a)–(c)** Out-of-plane (OP) hysteresis loops of normalized magnetization, $M/M_S$, for **(a)** $[Pt/CoB(2.4)/Ir]_4$, **(b)** $[Ir/CoB(2.0)/Pt]_4$ and **(c)** $[Pt/CoB(1.0)/MgO]_4$ multilayer stacks. Insets show the schematic overview of the multilayer stacks. **(d)–(o)** MFM images (scale bar: 1 μm) acquired over a range of *in situ* OP fields, after negative saturation, for **(d)–(g)** $[Pt/CoB(2.4)/Ir]_4$, **(h)–(k)** $[Ir/CoB(2.0)/Pt]_4$ and **(l)–(o)** $[Pt/CoB(1.0)/MgO]_4$ multilayer film. At $H = 0$, **(d), (h), (l)** show labyrinthine stripe textures. With increasing $H$, there is a gradual transition to skyrmion textures **(f), (j), (n)**.

$K_{eff}$ and low remanence, enabling the stabilization of multi-domain labyrinthine stripe states in the continuous film at zero applied field (Fig. 1(d), (h), (l)). Meanwhile, we demonstrate that in the presence of varying *in situ* OP fields, following negative OP saturation, skyrmions of average diameter ranging from 85 to 105, 98 to 106, and 105 to 152 nm (see S2) are stabilized across the Pt/CoB/Ir (Fig. 1(e)–(g)), Ir/CoB/Pt (Fig. 1(i)–(k)) and Pt/CoB/MgO (Fig. 1(m)–(o)) multilayers, respectively. Crucially, we highlight here the over 2× difference in CoB thicknesses across different multilayer stacks in displaying suitably low $K_{eff}$ for skyrmion stabilization. The paper henceforth focuses on addressing the origin of such magnetic property variations across these chiral multilayer platforms under as-deposited and annealed conditions. We examine the critical magnetic parameters (see equation (1)) $M_s$, $K_{eff}$ and $H_c$, derived from the $M(H)$ loops. Meanwhile, other magnetic parameters such as DMI and exchange are beyond the scope of this work.

First, we examine the magnetization dependence of CoB thickness across as-deposited multilayer stacks of Pt/CoB(2.0–3.6)/Ir, Ir/CoB(1.0–2.0)/Pt and Pt/CoB(1.0–2.0)/MgO — the selected CoB thickness ranges reflect the PMA to in-plane anisotropy (IPA) transition in all cases. Fig. 2(a)–(c) show the $M(H)$ hysteresis loops of representative multilayers with a fixed CoB thickness of 2 nm (see S3 for reference stack Pt/CoB/Pt). An apparent disparity in $M_s$ is observed across the as-deposited multilayer stacks, increasing across Pt/CoB/Ir < Ir/CoB/Pt < Pt/CoB/MgO (Fig. 2(d)–(f), Table 1). Meanwhile, a linear increase in $M_s.t$ is seen with increasing $t$, and the corresponding magnetic dead layer thickness ($t_d$) is derived from the $x$-axis intercept of the linear fit. The $t_d$ is non-zero across all stacks, albeit exhibiting opposite polarity — positive for Pt/CoB/Ir and negative for both Ir/CoB/Pt and Pt/CoB/MgO multilayers (Fig. 2(d)–(e), Table 1). Note that we expect the proximity-induced magnetization of Pt to contribute to the negative $t_d$ values, reported in recent work[18] and observed in our Pt/CoB/Pt stack (Table 1, see S3). Interestingly, the Pt/CoB/Pt $t_d$ is approximately twice that of the Ir/CoB/Pt stack, indicating the additive effect of comparable proximity-induced polarization of Pt sitting on either side of CoB. Strikingly, the $t_d$ becomes positive for the Pt/CoB/Ir, in contrast to the inverted stack where Pt sits atop of CoB. This underscores the presence of a thick CoB/Ir dead layer within the Pt/CoB/Ir stack[18] which compensates the Pt proximity-induced magnetization effects and gives rise to the lowest observed $M_s$ — a striking contrast to the Pt/CoB/Pt reference stack with the most substantial Pt polarization and highest $M_s$ (Table 1).

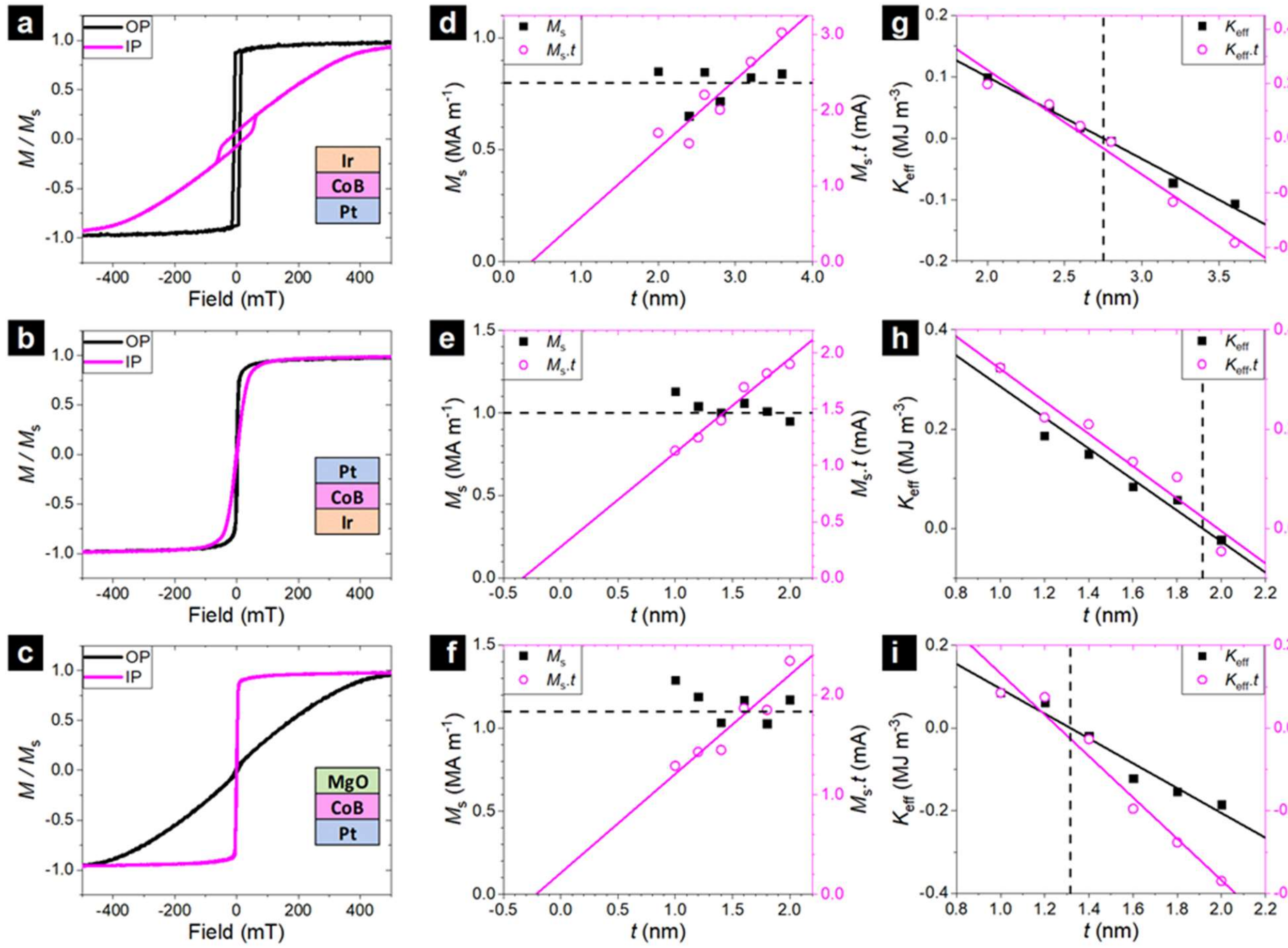

**Figure 2. Thickness dependence of magnetic parameters for as-deposited multilayer stacks. (a)–(c)** Out-of-plane (OP, black) and in-plane (IP, magenta) hysteresis loops of normalized magnetization, $M/M_S$, shown for representative samples of **(a)** Pt/CoB(2.0)/Ir, **(b)** Ir/CoB(2.0)/Pt and **(c)** Pt/CoB(2.0)/MgO. Insets show schematic representations of the multilayer stacks. **(d)–(i)** CoB thickness $t$ dependence of **(d)–(f)** saturation magnetization $M_S$ and $M_S.t$ and **(g)–(i)** effective magnetic anisotropy $K_{eff}$ and $K_{eff}.t$ for **(d), (g)** Pt/CoB(2.0–3.6)/Ir, **(e), (h)** Ir/CoB(1.0-2.0)/Pt and **(f), (i)** Pt/CoB(1.0–2.0)/MgO multilayer films. Solid lines show linear fits while dotted lines are guides to the eye for $M_s$ and $t_c$.

Next, we study the $K_{eff}$-dependence of CoB thickness across the as-deposited multilayer stacks (Fig. 2(g)–(i)). With increasing CoB thicknesses across multilayers, the $K_{eff}$ expectedly reduces and transits from PMA ($K_{eff} > 0$) to IPA ($K_{eff} < 0$). Importantly, the critical crossover CoB thickness ($t_c$) corresponding to the PMA-IPA transition (dashed black line in Figs. 2(g)–(i)) decreases across Pt/CoB/Ir > Ir/CoB/Pt > Pt/CoB/MgO (Table 1). The distinct variation in the $t_c$, as large as 2× across the symmetric reference (see S3) and asymmetric chiral stacks, reinforces the non-trivial role of dead layer(s) in these multilayers. Similar to the $K_{eff}$ dependence of $t$, the $K_{eff}.t$ dependence of $t$ show a monotonic decrease in the surface anisotropy ($K_{eff}.t$) with increasing CoB thicknesses, consistent with trends reported for Pt/Co/Ir and Ir/Co/Pt at Co thicknesses of more than 1 nm.[18] The surface anisotropy, as high as ~0.20 mJ $m^{-2}$ and ~0.32 mJ $m^{-2}$ for the Pt/CoB/Ir and Ir/CoB/Pt multilayers, respectively, is in the same order of magnitude as that reported for the Pt/Co/Ir and Ir/Co/Pt multilayers. Moreover, the surface anisotropy is higher for the FM/HM compared to FM/oxide (~0.09 mJ $m^{-2}$) stacks due to the stronger SOC contributions from both Pt/CoB and Ir/CoB in the former. In the next section, the origin of the stack structure dependent dead layer and its implications on the magnetic properties will be discussed.

**Table 1**: Summary of the saturation magnetization $M_S$, magnetic dead layer thickness $t_d$, and critical cross-over CoB thickness $t_c$, across the Pt/CoB/Ir, Ir/CoB/Pt, Pt/CoB/MgO and Pt/CoB/Pt multilayers.

| Stack | $M_s$ (MA $m^{-1}$) | $t_d$ (nm) | $t_c$ (nm) |
|---|---|---|---|
| Ta 3 / **Pt 5 / CoB 2–3.6 / Ir 1** / Pt 4 | 0.8 | 0.4 | 2.8 |
| Ta 3 / Pt 5 / **Ir 1 / CoB 1–2 / Pt 5** | 1.0 | -0.3 | 1.9 |
| Ta 3 / **Pt 5 / CoB 1–2 / MgO 1.5** / Pt 4 | 1.1 | -0.2 | 1.3 |
| Ta 3 / **Pt 5 / CoB 1-2 / Pt 5** (Reference sample, details in S3) | 1.2 | -0.6 | 2.3 |

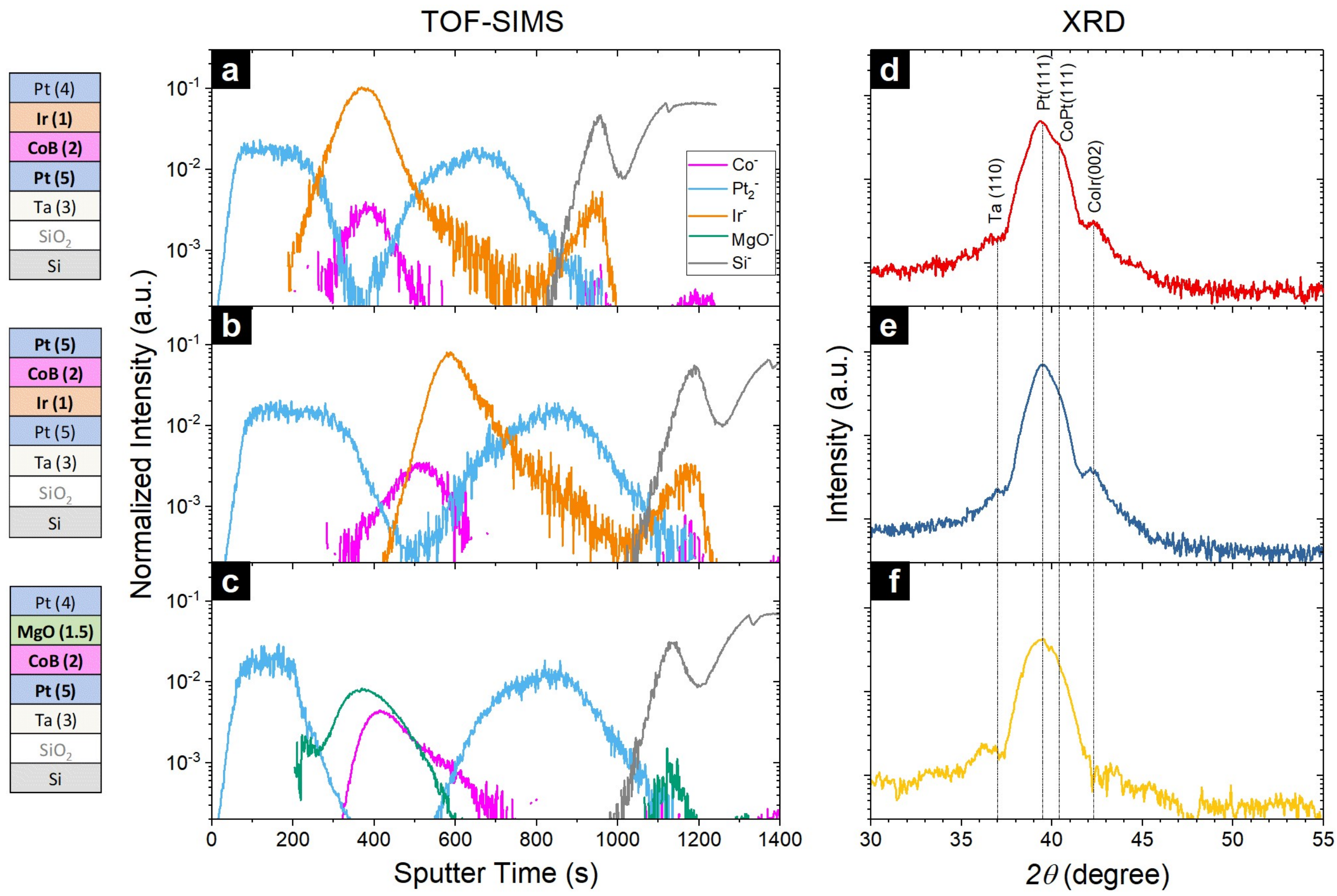

**Figure 3. TOF-SIMS and XRD analysis for as-deposited multilayer stacks. (a)–(c)** TOF-SIMS depth profiles showing secondary ion signals of Co (magenta), Pt (blue), Ir (orange), MgO (green), Si (grey) for **(a)** Pt/CoB(2.0)/Ir, **(b)** Ir/CoB(2.0)/Pt and **(c)** Pt/CoB(2.0)/MgO multilayer stacks. Increasing sputter time on the *x*-axis corresponds to increasing multilayer depth towards substrate. Ta secondary ion is excluded due to the low signal intensity in the negative polarity mode. **(d)–(f)** XRD $\theta$-$2\theta$ spectra for scan range of 30–55°, with step sizes of 0.05°, for **(d)** Pt/CoB(2.0)/Ir, **(e)** Ir/CoB(2.0)/Pt and **(f)** Pt/CoB(2.0)/MgO multilayer stacks. Left insets show schematic representations of the respective multilayer stacks, where nominal layer thicknesses in nm are indicated in parentheses.

## 4. Intermixing and Alloying Effects

TOF-SIMS depth profile characterization has been used in prior works to study intermixing effects in multilayer thin films.[32,33] To elucidate the origin of the magnetic dead layer in our stacks, we now turn to the TOF-SIMS analysis on representative as-deposited Pt/CoB(2.0)/Ir, Ir/CoB(2.0)/Pt and Pt/CoB(2.0)/MgO stacks, relative to a reference Pt/CoB/Pt stack (see S3). By utilizing TOF-SIMS analysis with $Cs^+$ ions for sputtering (see Methods), in-depth distribution of Ir, Co, Pt, MgO and Si secondary ions into the multilayers are monitored to provide a qualitative understanding of the interlayer intermixing across the stacks (Fig. 3(a)–(c)). Interestingly, distinctive intermixing behaviour is observed for the Pt/CoB/Ir and Ir/CoB/Pt stacks (Fig. 3(a)–(b)), suggesting a strong influence of the inversion of stack ordering on the extent of intermixing. For Pt/CoB/Ir (Fig. 3(a)), the top Ir layer overlaps in entirety with the bottom Co layer, presenting a stark contrast to the misaligned top CoB and bottom Ir signatures for the Ir/CoB/Pt (Fig. 3(b)). The greater extent of intermixing in the former may be attributed to the high surface mobility of Ir adatoms with larger sputtering power (Ir: 0.3 kW, Pt: 0.075 kW).[34,35] We posit that the larger sputtering power-induced intermixing may result in enhanced Co-Ir alloying.[34,40] Meanwhile, similar Co and Pt intermixing profiles are observed on both Pt/CoB and CoB/Pt interfaces across Pt/CoB/Ir, Ir/CoB/Pt (Fig. 3(a), (b)) and Pt/CoB/Pt (see S3) stacks, suggestive of comparable intermixing behaviour and proximity-induced magnetization effects of Pt deposited on either side of CoB. Similar to the Ir/CoB/Pt stack, the Pt/CoB/MgO stack (Fig. 3(c)) exhibits a relatively misaligned CoB and MgO intensity profile, consistent with much reduced intermixing between the layers.

Next, we examine the crystallographic textural variation across the as-deposited stacks (Fig. 3(d)–(f)). The Bragg peak positions at 37.0° and 39.5° are associated with the body-centred cubic (110) and face-centred cubic (*fcc*) (111) textures of the Ta and Pt underlayers, respectively (See S4). In both Pt/CoB/Ir and Ir/CoB/Pt stacks, Co-Ir hexagonal close packed (*hcp*) (002) peak at ~42.5° are observed, reflecting the presence of Co-Ir alloying (Fig. 3(d)–(e)). The formation of Co-Ir alloy understandably results in the leftward shift of the pure Co (002) Bragg peak ($2\theta$: 44.5°) due to a larger lattice constant induced by the bigger Ir radius.[36,37] Additionally, there are faint signs of a weak shoulder peak at ~40.8° attributed to slight Co-Pt *fcc* (111) formation. Further, the absence of any significant Co-based Bragg peaks in the $2\theta$ scan for the as-deposited Pt/CoB/MgO stack (Fig. 3(f)) suggests that the film is predominantly amorphous in nature (see S5).

Evidently, the lower $M_s$ observed for the metallic stacks (Fig. 2(d)–(f), Table 1) may be ascribed to the formation of Co-Ir alloy (Fig. 3(d),(e)). Consistent with earlier work, $M_s$ reduces monotonically with increasing Ir content in Co-Ir alloy,[21,36,38] to as low as 0.7 MA m$^{-1}$ with Ir content of 22%.[21] The greater reduction in $M_s$ for the Pt/CoB/Ir stack, relative to the inverted stack, arises from the more significant CoB-Ir intermixing when Ir sits atop of CoB (Fig. 3(d)–(e)).[18] Further, the presence of the thicker CoIr magnetic dead layer in the Pt/CoB/Ir stack compensates the Pt polarization effects, resulting in its positive $t_d$ value. Importantly, substantial intermixing in the Pt/CoB/Ir multilayer reduces the effective CoB thickness and increases interfacial roughness,[18,22] accounting for the observation of larger $t_c$ required for PMA-IPA transition (Section 3, Fig. 2).

## 5. Thermal Annealing Effects

Thermal annealing of multilayer stacks brings about considerable changes to the interfacial roughness, interlayer atomic diffusion, crystallization, phase transitions, hence altering their magnetic and transport properties.[20,23,39–42] Moderate annealing, e.g. over 100–300 °C, is expected to limit the detrimental effects of thermally activated intermixing due to diffusion, while enabling improved material parameters relevant to device performance such as anisotropy, magnetoresistance and resistance-area.[22,43] Here, we study the thermal annealing temperature ($T$) dependence of magnetic properties across the three multilayer stacks at a fixed CoB thickness of 2 nm (Fig. 4). There is a marginal variation of $M_s$ across the stacks (Fig. 4(a)). With increasing $T$ up to 300 °C, the $K_{eff}$ increases monotonically by ~0.1–0.3 MJ m$^{-3}$ for the three stacks. (Fig. 4(b)). On a related note, the $H_c$ also increases markedly with higher $T$ across the metallic stacks, with an enhancement as high as 13× (~1.6 to 21.6 mT) for Ir/CoB/Pt (Fig. 4(c)).

TOF-SIMS depth profile analysis in Fig. 5(a)–(c) shows the preferential inward migration of the left and right Pt peaks (blue) upon annealing at $T$ = 300 °C across all samples, suggesting the in-diffusion of Pt atoms from both the top layer and underlayer towards the middle Ir/CoB and CoB/MgO layers. This may be understood as arising from the diffusion rate ($J$) dependence of the atomic concentration gradient based on Fick's first Law[44]

$$J = -D\frac{\delta C}{\delta x} \tag{3}$$

where $D$ is the diffusion coefficient, $C$ is the concentration and $x$ is the local position within the stack. Expectedly, the thicker Pt layers, up to 5× thicker than other multilayer components, possess higher thermally-driven diffusion kinetics.[45] Meanwhile, the simultaneous reduction in the Co peak (magenta) intensity across all samples, along with the in-diffusion of Pt, may be attributed to the higher diffusivity of the small (125 pm) and light (8.9 g cm$^{-3}$) Co atoms — 1.5× smaller and 2.5× lighter with respect to Pt and Ir atoms. Notably, the intermixing effect is particularly pronounced for the metallic stacks wherein the two Pt peaks seem to merge with a discernible increase in Pt intensity mid-way (Fig. 5(a)–(b)). On the contrary, the Pt intermixing is less extensive in the CoB/MgO stack where the oxide layer is an effective barrier against thermal-induced Pt migration (Fig. 5(c)). Such disparity in intermixing behaviours highlights the better diffusivity of Pt through polycrystalline Ir via lattice defects, particularly grain boundaries,[34,46] compared to metal-oxides known to be good diffusion barriers.[47,48]

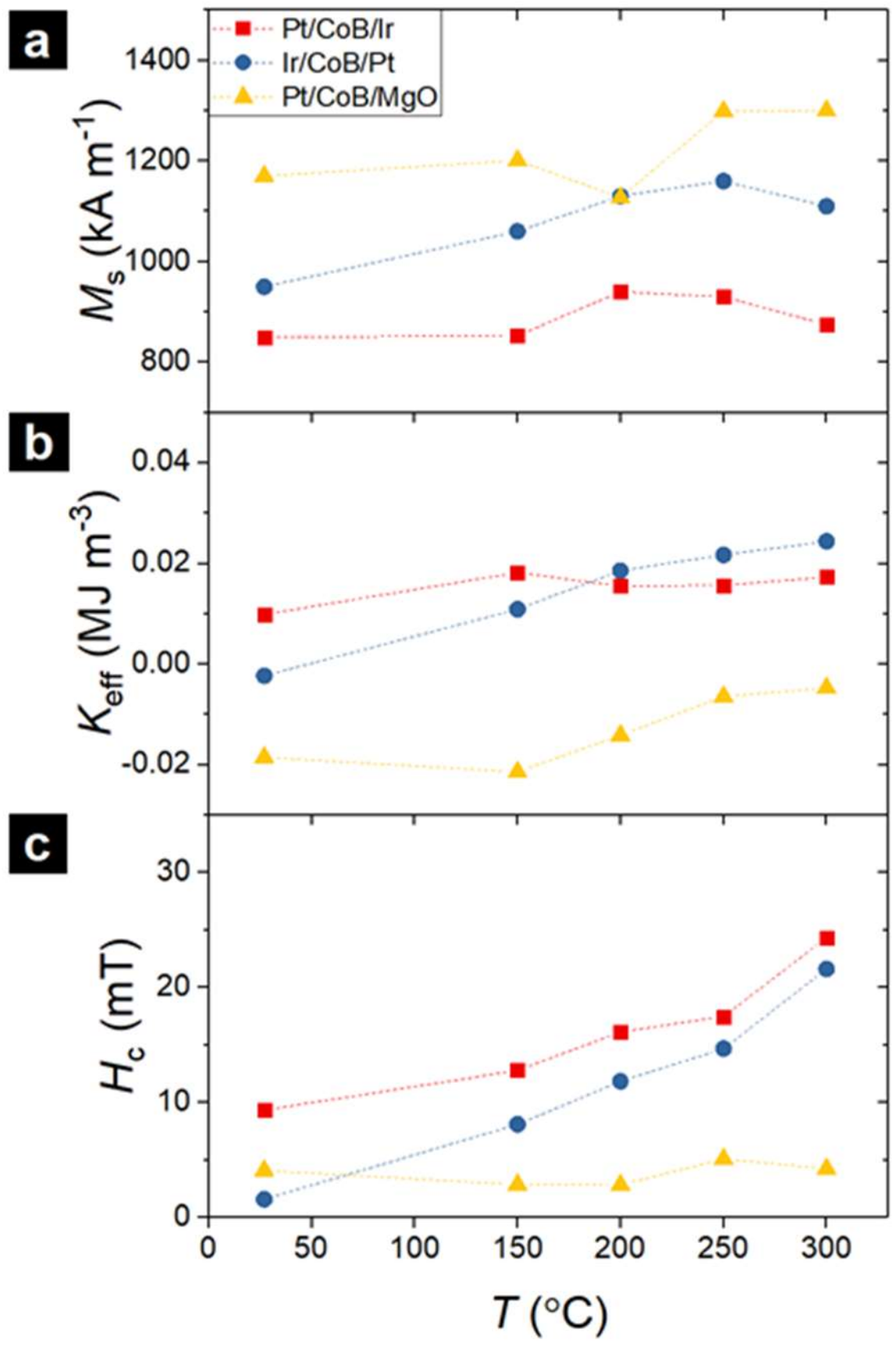

**Figure 4. Annealing temperature dependence of magnetic parameters across multilayer stacks.** Annealing temperature $T$ dependence of **(a)** $M_s$, **(b)** $K_{eff}$ and **(c)** coercivity ($H_c$) across Pt/CoB(2.0)/Ir, Ir/CoB(2.0)/Pt and Pt/CoB(2.0)/MgO stacks.

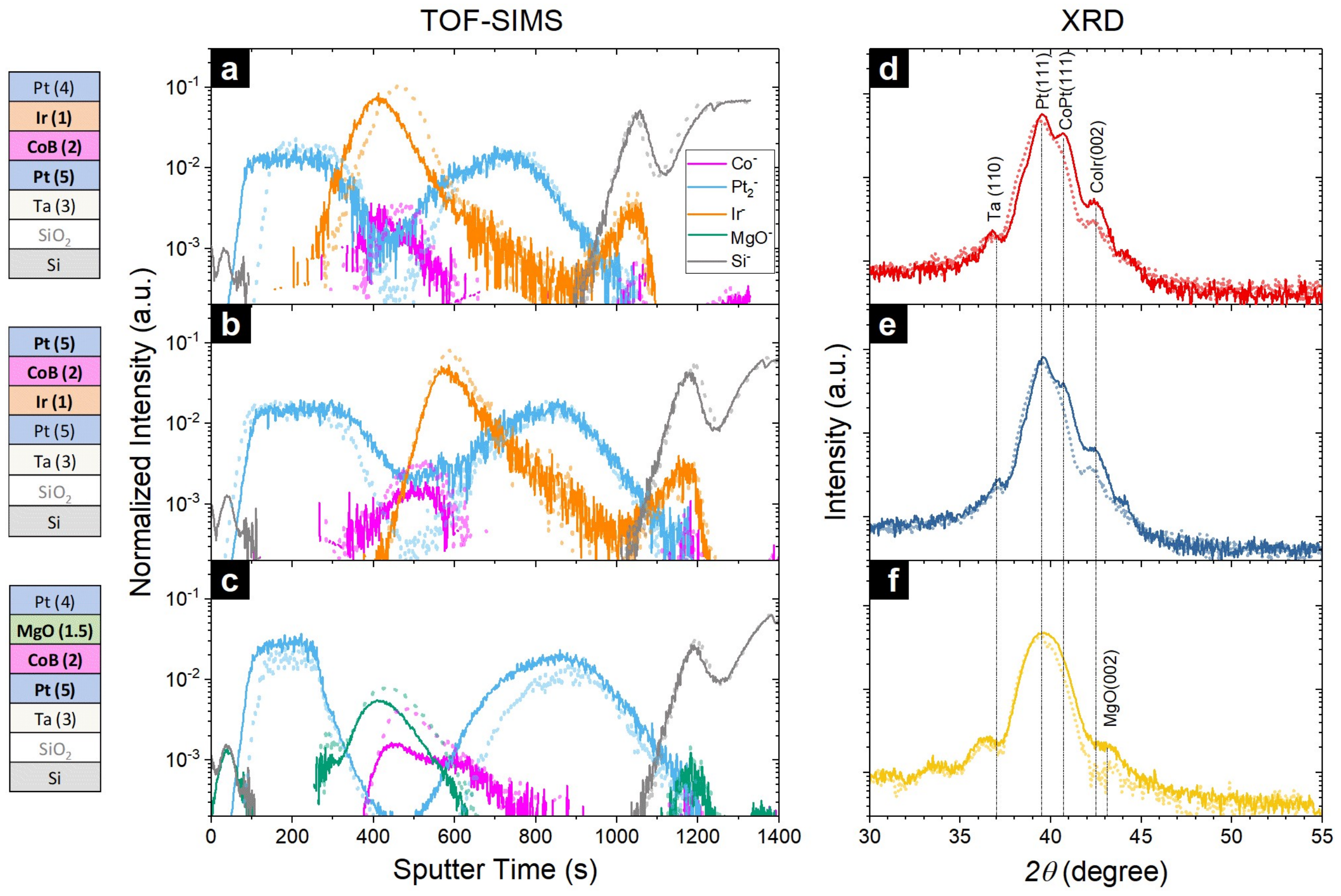

**Figure 5. TOF-SIMS and XRD analysis across annealed multilayer stacks. (a)–(c)** TOF-SIMS depth profiles showing secondary ion signals of Co (magenta), Pt (blue), Ir (orange), MgO (green), Si (grey) for **(a)** Pt/CoB(2.0)/Ir, **(b)** Ir/CoB(2.0)/Pt and **(c)** Pt/CoB(2.0)/MgO stacks when annealed at 300 °C (solid line), compared to the as-deposited profiles (dotted line, see Fig. 3). Increasing sputter time on the *x*-axis corresponds to increasing multilayer depth towards substrate. For ease of comparison within each stack, the *x*-axis of the annealed and as-deposited depth profiles are aligned with respect to the $SiO_2$ substrate. Ta secondary ion is excluded due to the low signal intensity in the negative polarity mode. **(d)–(f)** XRD $\theta$-$2\theta$ spectra for scan range of 30–55°, with step sizes of 0.05°, for **(d)** Pt/CoB(2.0)/Ir, **(e)** Ir/CoB(2.0)/Pt and **(f)** Pt/CoB(2.0)/MgO multilayer stacks when as-deposited (dotted line, Fig. 3) and annealed at 300 °C (solid line). Left insets show schematic representations of the respective multilayer stacks, where nominal layer thicknesses in nm are indicated in parentheses.

In both as-deposited Pt/CoB/Ir and Ir/CoB/Pt stacks (Fig. 3(d)–(e)), Co-Ir *hcp* (002) peak at ~42.5° are observed. Upon annealing at $T$ = 300 °C, the intensity of the Co-Ir *hcp* (002) peak increases, indicating the introduction of additional thermal-induced CoIr alloying. Importantly, we observed the emergence of fundamental Co-Pt *fcc* (111) peak at ~40.8° across both metallic multilayers,[20,39,40] evidently a result of the inward diffusion of Pt into the CoB layer (Fig. 5(a)–(b)). The Co-Pt peak indicates the likely formation of $CoPt_3$ binary alloy, wherein a Co-Pt *fcc* (111) peak position of 40.5° has been reported for Pt content up to 70%.[20] The Co-Pt crystallite size is estimated to be ~12 nm in both metallic multilayers. The formation of $CoPt_3$ crystalline phase with increasing $T$ introduces extra magnetocrystalline anisotropy contribution, thereby enhancing the $K_{eff}$ and $H_c$.[20,40,49] Additionally, the limited variation in $M_s$ across $T$ is consistent with the near-constant $M_s$ reported for temperature-dependent CoPt formation.[20]

Upon annealing the Pt/CoB/MgO stack at 300 °C, we observe a broad and weak peak at ~43.1° (Fig. 5(f)), which is suggestive of the formation of crystalline MgO (200). Such crystallization may be driven by the diffusion of Boron out of CoB into MgO to form a Boron enriched mixed layer of Mg-B-O. Similar crystallization of MgO has been reported for the CoFeB/MgO system at moderate annealing temperatures of 300 °C.[23] Notably, we do not observe any Co peaks in post-annealed films, indicating the absence of a discernible polycrystalline Co layer, e.g. at the CoB/MgO interface upon annealing at moderate temperatures. This is in line with previous work on CoFeB/MgO systems, indicating the onset of CoFe crystallization occurs only at high temperatures > 300 °C.[41] Further, this suggests negligible thermal-induced CoPt alloying arising from the relatively mild Pt intermixing (Fig. 5(c)) that is impeded by the effective oxide barrier. Here, the PMA enhancement with annealing (Fig. 4(b)) is consistent with earlier reports on annealed Pt/Co/Si, Al, Mg-based oxides,[50,51] and ascribed to the hybridization of Co-3*d* with O-

2*p* orbitals with increasing oxygen content at the CoB/MgO interface.[42]

## 6. Discussion

In summary, we have presented a comprehensive study of detailing the intermixing and thermal annealing dependence of magnetic parameters across amorphous CoB-based multilayer films — Pt/CoB/Ir, Ir/CoB/Pt and Pt/CoB/MgO — hosting RT chiral magnetic textures. First, prevailing intermixing phenomenon in the Pt/CoB/Ir and Ir/CoB/Pt multilayers result in the formation of the magnetic dead layer of CoIr alloy, accounting for the overall low $M_s$, high $t_d$ and high $t_c$ in the metallic multilayers. Across the metallic stacks, the stack ordering of Ir atop CoB gives rise to increased intermixing, and correspondingly the lowest $M_s$, highest $t_d$ and highest $t_c$. Second, discernible formation of CoPt alloy and crystallization of MgO are observed upon annealing the metallic and oxide stacks, resulting in an enhancement of $K_{eff}$. Lastly, we demonstrate large PMA in the Pt/CoB/Ir and Ir/CoB/Pt stacks which can be modulated with CoB thickness and annealing temperature variations. Crucially, the amorphous nature of CoB is retained across as-deposited and annealed Pt/CoB/MgO stacks.

The development of an array of amorphous CoB-based multilayers hosting RT chiral magnetic textures paves the way for imminent material and device directions. While the metallic Pt/CoB/Ir and Ir/CoB/Pt stacks provide large tunable PMA appropriate for a range of device applications, the Pt/CoB/MgO multilayer is well-suited for tunnelling magnetoresistance readout of the magnetic textures in magnetic tunnel junction devices. In this vein, material design efforts built upon a hybrid multilayer stack — integrating both the metallic and oxide multilayer systems — are of immediate relevance for the exploration of memory and neuromorphic computing devices. Finally, insights from our work on the influence of intermixing, arising from materials stack design and annealing, sets the foundation for tailoring magnetic interactions to cater to a range of technological applications. Future studies should be geared towards elucidating the effects of intermixing on other critical magnetic interactions such as DMI and *A* to further fine-tune chiral texture parameters such as stability, size and density.

## Acknowledgements

We acknowledge Sze Ter Lim and S. N. Piramanayagam for insightful discussions, and Huiqing Xie for performing x-ray photoelectron spectroscopy analysis on the CoB film composition. This work was supported by the SpOT-LITE programme (Grant Nos. A18A6b0057), funded by Singapore's RIE2020 initiatives.

[†]anjan@imre.a-star.edu.sg and hopin@imre.a-star.edu.sg

## References

1. Emori, S., Bauer, U., Ahn, S. M., Martinez, E. & Beach, G. S. D. Current-driven dynamics of chiral ferromagnetic domain walls. *Nat. Mater.* **12**, 611–616 (2013).
2. Parkin, S. & Yang, S. H. Memory on the racetrack. *Nat. Nanotechnol.* **10**, 195–198 (2015).
3. Dupé, B., Bihlmayer, G., Böttcher, M., Blügel, S. & Heinze, S. Engineering skyrmions in transition-metal multilayers for spintronics. *Nat. Commun.* **7**, 11779 (2016).
4. Nagaosa, N. & Tokura, Y. Topological properties and dynamics of magnetic skyrmions. *Nat. Nanotechnol.* **8**, 899–911 (2013).
5. Soumyanarayanan, A., Reyren, N., Fert, A. & Panagopoulos, C. Emergent phenomena induced by spin–orbit coupling at surfaces and interfaces. *Nature* **539**, 509–517 (2016).
6. Soumyanarayanan, A. *et al.* Tunable Room Temperature Magnetic Skyrmions in Ir/Fe/Co/Pt Multilayers. *Nat. Mater.* **16**, 898–904 (2017).
7. Ho, P. *et al.* Geometrically Tailored Skyrmions at Zero Magnetic Field in Multilayered Nanostructures. *Phys. Rev. Appl.* **11**, 024064 (2019).
8. Chen, G. *et al.* Out-of-plane chiral domain wall spin-structures in ultrathin in-plane magnets. *Nat. Commun.* **8**, 15302 (2017).
9. Legrand, W. *et al.* Hybrid chiral domain walls and skyrmions in magnetic multilayers. *Sci. Adv.* **4**, eaat0415 (2018).
10. Blundell, S. *Magnetism in Condensed Matter*. (Oxford Univeristy Press Inc., 2001).
11. Johnson, M. T., Bloemen, J. H., Den Broeder, J. A. & De Vries, J. J. Magnetic anisotropy in metallic multilayers. *Rep. Prog. Phys* **59**, 1409–1458 (1996).
12. Ikeda, S. *et al.* A perpendicular-anisotropy CoFeB-MgO magnetic tunnel junction. *Nat. Mater.* **9**, 721–724 (2010).
13. Nakajima, N. *et al.* Perpendicular Magnetic Anisotropy Caused by Interfacial Hybridization via Enhanced Orbital Moment in Co/Pt Multilayers: Magnetic Circular X-Ray Dichroism Study. *Phys. Rev. Lett.* **81**, 5229–5232 (1998).
14. Khoo, K. H. *et al.* First-principles study of perpendicular magnetic anisotropy in CoFe/MgO and CoFe/Mg3B2O6 interfaces. *Phys. Rev. B* **87**, 174403 (2013).
15. Peng, S. *et al.* Origin of interfacial perpendicular magnetic anisotropy in MgO/CoFe/metallic capping layer structures. *Sci. Rep.* **5**, 18173 (2015).
16. Carcia, P. F. Perpendicular magnetic anisotropy in Pd/Co and Pt/Co thin-film layered structures. *J. Appl. Phys.* **63**, 5066–5073 (1988).
17. Kubota, H. *et al.* Enhancement of perpendicular magnetic anisotropy in FeB free layers using a thin MgO cap layer. *J. Appl. Phys.* **111**, 07C723 (2012).
18. Lau, Y.-C. *et al.* Giant perpendicular magnetic anisotropy in Ir/Co/Pt multilayers. *Phys. Rev. Mater.* **3**, 104419 (2019).
19. Gweon, H. K., Yun, S. J. & Lim, S. H. A very large perpendicular magnetic anisotropy in Pt/Co/MgO trilayers fabricated by controlling the MgO sputtering power and its thickness. *Sci. Rep.* **8**, 1266 (2018).
20. Makarov, D. *et al.* Nonepitaxially grown nanopatterned Co–Pt alloys with out-of-plane magnetic anisotropy. *J. Appl. Phys.* **106**, 114322 (2009).
21. Hashimoto, A., Saito, S. & Takahashi, M. A soft magnetic underlayer with negative uniaxial magnetocrystalline anisotropy for suppression of spike noise and wide adjacent track erasure in perpendicular recording media. *J. Appl. Phys.* **99**, 08Q907 (2006).
22. Carlomagno, I. *et al.* Co-Ir interface alloying induced by thermal annealing. *J. Appl. Phys.* **120**, 195302 (2016).
23. Mukherjee, S. *et al.* Role of boron diffusion in CoFeB/MgO

magnetic tunnel junctions. *Phys. Rev. B* **91**, 085311 (2015).
24. Woo, S. *et al.* Observation of room-temperature magnetic skyrmions and their current-driven dynamics in ultrathin metallic ferromagnets. *Nat. Mater.* **15**, 501 (2016).
25. Juge, R. *et al.* Current-Driven Skyrmion Dynamics and Drive-Dependent Skyrmion Hall Effect in an Ultrathin Film. *Phys. Rev. Appl.* **12**, 044007 (2019).
26. Zeissler, K. *et al.* Diameter-independent skyrmion Hall angle observed in chiral magnetic multilayers. *Nat. Commun.* **11**, 428 (2020).
27. Xu, Y. B., Greig, D., Mitchell, A. L., Seddon, E. A. & Matthew, J. A. D. The spin-dependent electronic structure of amorphous magnetic alloys. *J. Appl. Phys.* **81**, 4419–4421 (1997).
28. Tanaka, H. *et al.* Electronic structure and magnetism of amorphous Co1−xBx alloys. *Phys. Rev. B* **47**, 2671–2677 (1993).
29. Lordan, D., Wei, G., McCloskey, P., O'Mathuna, C. & Masood, A. Origin of perpendicular magnetic anisotropy in amorphous thin films. *Sci. Rep.* **11**, 3734 (2021).
30. Langford, J. I. & Wilson, A. J. C. Scherrer after sixty years: A survey and some new results in the determination of crystallite size. *J. Appl. Crystallogr.* **11**, 102–113 (1978).
31. Boulle, O. *et al.* Room-temperature chiral magnetic skyrmions in ultrathin magnetic nanostructures. *Nat. Nanotechnol.* **11**, 449–454 (2016).
32. Le Guen, K. *et al.* Observation of an asymmetrical effect when introducing Zr in Mg/Co multilayers. *Appl. Phys. Lett.* **98**, 251909 (2011).
33. Samardak, A. S. *et al.* Enhancement of perpendicular magnetic anisotropy and Dzyaloshinskii–Moriya interaction in thin ferromagnetic films by atomic-scale modulation of interfaces. *NPG Asia Mater.* **12**, 51 (2020).
34. Aboulfadl, H., Gallino, I., Busch, R. & Mücklich, F. Atomic scale analysis of phase formation and diffusion kinetics in Ag/Al multilayer thin films. *J. Appl. Phys.* **120**, 195306 (2016).
35. Harsha, K. S. S. *Principles of Vapor Deposition of Thin Films*. (Elsevier, 2006).
36. Xu, F., Zhang, S., Yang, D., Wang, T. & Li, F. High-frequency properties of oriented hcp-Co 1−x Ir x (0.06 ≤ x ≤ 0.24) soft magnetic films. *J. Appl. Phys.* **117**, 17B725 (2015).
37. Zhang, S. *et al.* First-principles study of the easy-plane magnetocrystalline anisotropy in bulk hcp Co 1− x Ir x. *J. Appl. Phys.* **126**, 083907 (2019).
38. Jiao, J., Wang, T., Ma, T., Wang, Y. & Li, F. Achievement of Diverse Domain Structures in Soft Magnetic Thin Film through Adjusting Intrinsic Magnetocrystalline Anisotropy. *Nanoscale Res. Lett.* **12**, 21 (2017).
39. Trung, T. T., Nhung, D. T., Nam, N. H. & Luong, N. H. Synthesis and Magnetic Properties of CoPt Nanoparticles. *J. Electron. Mater.* **45**, 3621–3623 (2016).
40. Sun, X. *et al.* Synthesis and magnetic properties of CoPt nanoparticles. *J. Appl. Phys.* **95**, 6747–6749 (2004).
41. Swamy, G. V. *et al.* Effect of thermal annealing on Boron diffusion, micro-structural, electrical and magnetic properties of laser ablated CoFeB thin films. *AIP Adv.* **3**, 072129 (2013).
42. Yang, H. X. *et al.* First-principles investigation of the very large perpendicular magnetic anisotropy at Fe|MgO and Co|MgO interfaces. *Phys. Rev. B* **84**, 054401 (2011).
43. Sousa, R. C. *et al.* Large tunneling magnetoresistance enhancement by thermal anneal. *Appl. Phys. Lett.* **73**, 3288–3290 (1998).
44. Kittel, C. *Introduction to Solid State Physics*. (John Wiley & Sons Inc., 2005).
45. Cialone, M. *et al.* Tailoring magnetic properties of multicomponent layered structure via current annealing in FePd thin films. *Sci. Rep.* **7**, 16691 (2017).
46. Belova, I. V. & Murch, G. E. Diffusion in nanocrystalline materials. *J. Phys. Chem. Solids* **64**, 873–878 (2003).
47. Ahamad Mohiddon, M. *et al.* Chromium oxide as a metal diffusion barrier layer: An x-ray absorption fine structure spectroscopy study. *J. Appl. Phys.* **115**, 044315 (2014).
48. Harper, J. M. E., Hörnström, S. E., Thomas, O., Charai, A. & Krusin-Elbaum, L. Mechanisms for success or failure of diffusion barriers between aluminum and silicon. *J. Vac. Sci. Technol. A Vacuum, Surfaces, Film.* **7**, 875–880 (1989).
49. Ricardo-Chávez, J. L., Muñoz-Navia, M. & Ruiz-Díaz, P. Magnetocrystalline anisotropy of small CoPt binary alloy metal clusters: interplay between structure, chemical composition, and spin-orbit coupling. *J. Nanoparticle Res.* **22**, 215 (2020).
50. Rodmacq, B., Manchon, A., Ducruet, C., Auffret, S. & Dieny, B. Influence of thermal annealing on the perpendicular magnetic anisotropy of Pt/Co/AlOx trilayers. *Phys. Rev. B* **79**, 024423 (2009).
51. Nistor, L. E., Rodmacq, B., Auffret, S. & Dieny, B. Pt/Co/oxide and oxide/Co/Pt electrodes for perpendicular magnetic tunnel junctions. *Appl. Phys. Lett.* **94**, 012512 (2009).

## Supporting Information for

# Intermixing Induced Anisotropy Variations in CoB-based Chiral Multilayer Films

## S1. CoB Composition

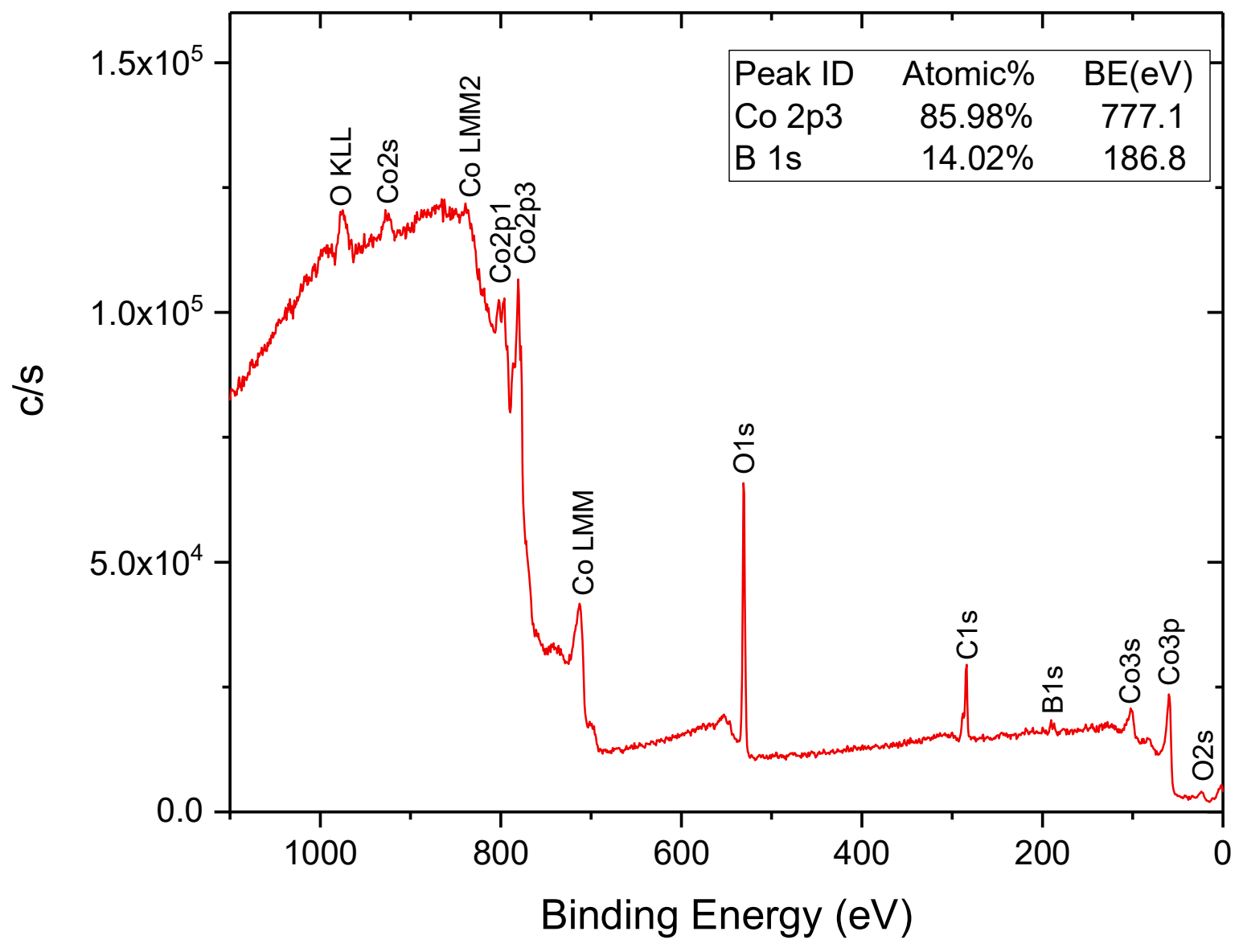

**Figure S1. X-ray photoelectron spectroscopy**. XPS surface compositional analysis on Si/$SiO_2$/CoB(20) (thickness in nm in parentheses) thin film.

**X-ray Photoelectron Spectroscopy (XPS) analysis.** A $Co_{80}B_{20}$ composite target was utilized for the magnetron sputtering deposition of the CoB thin film. To ascertain the chemical composition of the sputtered CoB thin film, XPS surface compositional analysis was performed, using the Quantera SXM$^{TM}$, on a 20nm-thick CoB film deposited on thermally oxidized Si wafer. Slight contaminants of C and O elements are detected on the surface, while an average composition of $Co_{86}B_{14}$ is determined from the Co and B elements detected.

## S2. Skyrmion Sizes

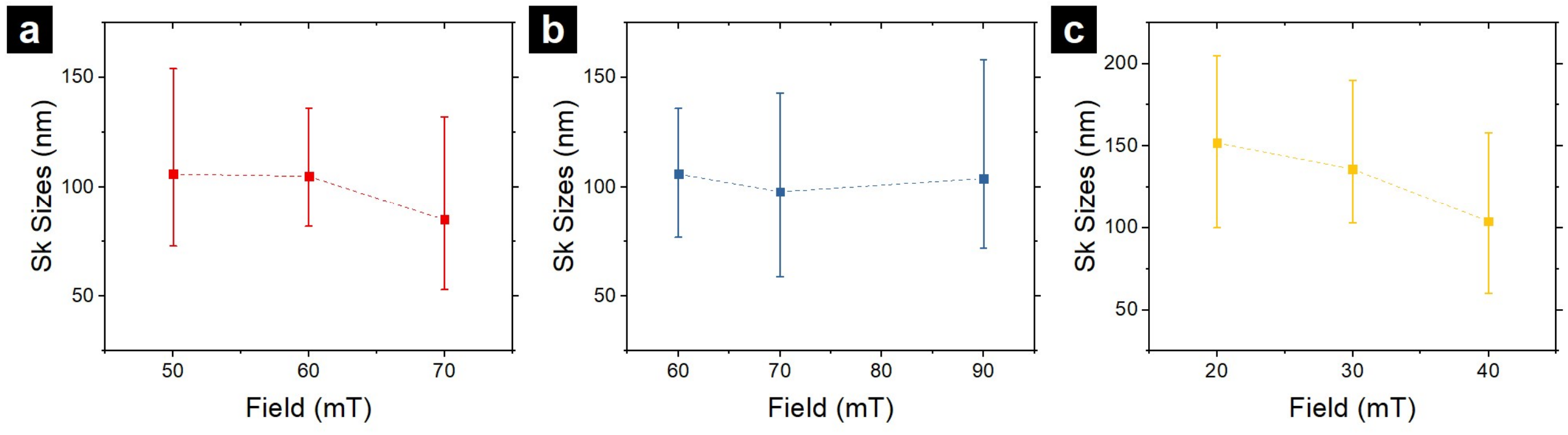

**Figure S2. Skyrmion sizes. (a)–(c)** Skyrmion sizes obtained by examining magnetic force microscopy images of **(a)** [Pt/CoB(2.4)/Ir]4, **(b)** [Ir/CoB(2.0)/Pt]4 and **(c)** [Pt/CoB(1.0)/MgO]4 multilayer stacks (nominal layer thicknesses in nm in parentheses) for *in situ* out-of-plane (OP) fields of 50–70, 60–90 and 20–40 mT, respectively, following negative saturation (detailed in Fig. 1(e)–(g), (i)–(k), (m)–(o)).

**Skyrmion sizes**. Manuscript Fig. 1(d)–(o) details the evolution of magnetic states acquired over a range of applied *in situ* out-of-plane (OP) fields ($H$), following negative saturation, across the multilayers. At applied $H = 0$ mT, the domain periodicity for the labyrinth stripe texture, obtained from the Fourier transform peak of the zero-field magnetic force microscopy (MFM) images, are 136, 121 and 228 nm for Pt/CoB/Ir, Ir/CoB/Pt and Pt/CoB/MgO multilayers respectively. With increasing *in situ* OP fields, there is a monotonic reduction in the skyrmion sizes from 105 to 85, and 152 to 105 nm for Pt/CoB/Ir and Pt/CoB/MgO multilayers, respectively; while a near-constant variation of ~98–106 nm for Ir/CoB/Pt multilayer. The skyrmion sizes are determined from the full width at half maximum of the Gaussian fit to the line cut of the skyrmions identified on the MFM.

## S3. Pt/CoB/Pt Reference Multilayer Stack

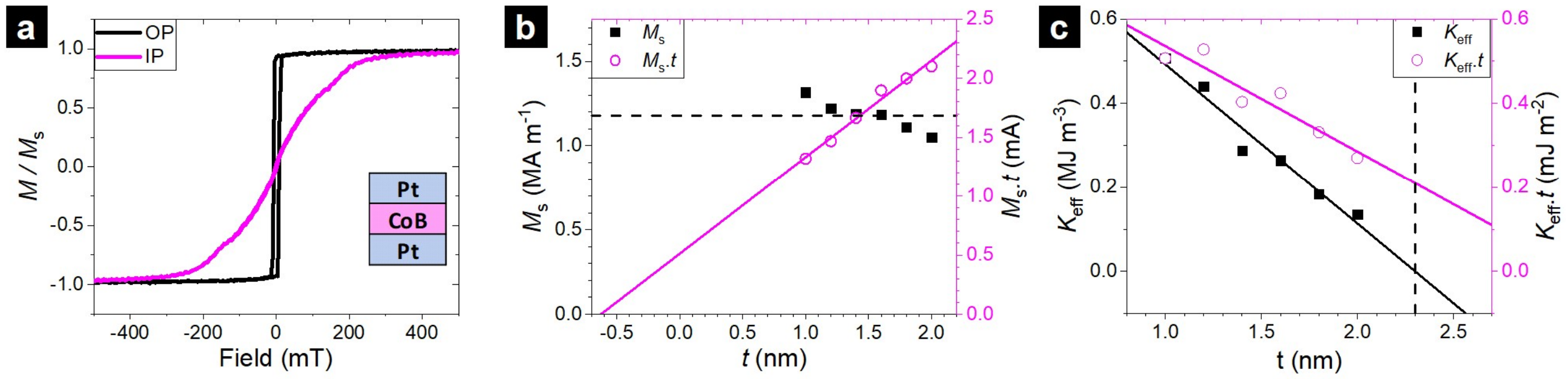

**Figure S3. Thickness dependence of magnetic parameters for as-deposited Pt/Co/Pt multilayer stack. (a)** Out-of-plane (OP, black) and in-plane (IP, magenta) hysteresis loops of normalized magnetization, $M/M_s$, **(b)** CoB thickness $t$ dependence of saturation magnetization $M_s$ and $M_s.t$ and **(c)** effective magnetic anisotropy $K_{eff}$ and $K_{eff}.t$ for the representative sample Pt/Co(2.0)/Pt. Solid lines show linear fits while dotted lines are guides to the eye for $M_s$ and $t_c$.

A reference multilayer stack comprising Ta(3)/Pt(5)/CoB(1-2)/Pt(5) (nominal layer thicknesses in nm in parentheses) was deposited on pre-cleaned thermally oxidized Si coupon. Fig. S3 shows the magnetization hysteresis loop $M(H)$ of a representative Pt/CoB(2.0)/Pt stack with a fixed CoB thickness of 2 nm, similar to the other two metallic multilayer stacks – Pt/CoB(2.0)/Ir and Ir/CoB(2.0)/Pt (manuscript Fig. 2). Across the three metallic stacks, Pt/CoB/Pt displays the highest saturation magnetization, $M_s$, (1.2 MA $m^{-1}$) and the most negative magnetic dead layer thickness, $t_d$, (-0.6 nm) (manuscript Table 1). This is ascribed to the highest Pt polarization arising from the proximity-induced magnetization effects from both the Pt-CoB interfaces. Interestingly, the Pt/CoB/Pt $t_d$ is approximately twice that of the Ir/CoB/Pt stacks, indicating the additive effect of comparable proximity-induced polarization of Pt sitting on either side of CoB. Additionally, the critical crossover CoB thickness ($t_c$) of the Pt/CoB/Pt stack (2.3 nm) falls midway between the Pt/CoB/Ir (2.8 nm) and Ir/CoB/Pt (1.9 nm), indicating intermediate intermixing and moderate magnetic dead layer effects in the Pt/CoB/Pt stacks (see Fig. S4 and S5). Further, we observed surface anisotropy as high as ~0.60 mJ $m^{-2}$ in the Pt/CoB/Pt stack — 3× more than the Pt/CoB/Ir and Ir/CoB/Pt metallic stacks, possibly due to stronger spin-orbit coupling contributions from both Pt/CoB interfaces.

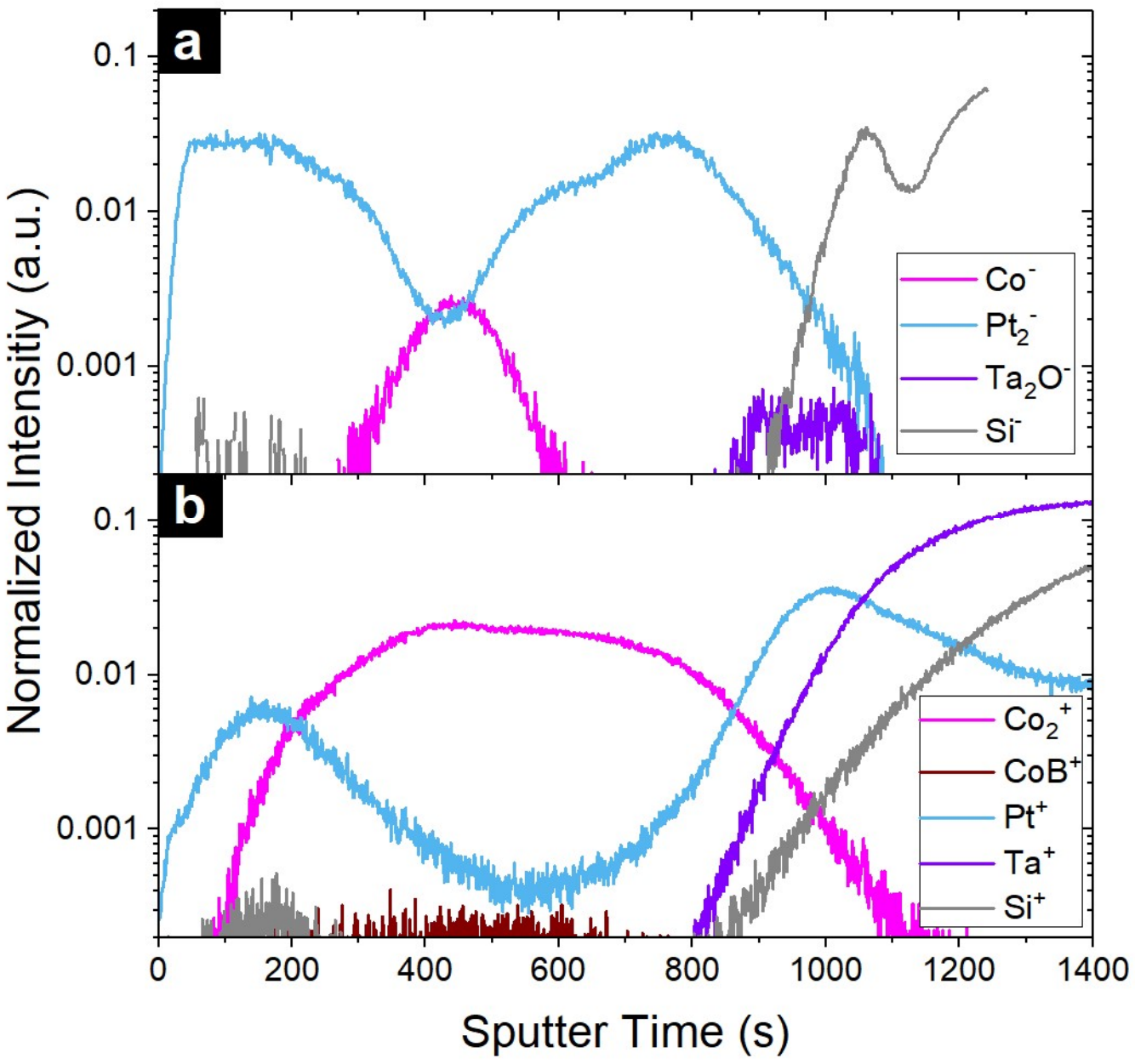

**Figure S4. TOF-SIMS analysis for as-deposited reference Pt/CoB/Pt**. **(a)–(b)** Time-of-flight secondary ion mass spectrometry (TOF-SIMS) depth profiles for Ta(3)/Pt(5)/CoB(2)/Pt(5) multilayer (thickness in nm in parentheses) using **(a)** negative polarity mode with $Cs^+$ sputter ion source showing secondary ion signals of Co (magenta), Pt (blue), Ta (violet) and Si (grey) and **(b)** positive polarity mode with $O_2^+$ sputter ion source showing secondary ion signals of Co (magenta), Pt (blue), CoB (wine), Ta (violet) and Si (grey). Increasing sputter time on the *x*-axis corresponds to increasing multilayer depth towards substrate.

**TOF-SIMS Depth Profile.** By utilizing TOF-SIMS analysis with $Cs^+$ ions (negative polarity mode) and $O_2^+$ ions (positive polarity mode) for sputtering, the profiles of Co, Pt, CoB, Ta and Si are mapped for the Pt/CoB/Pt reference sample. The Ta layer produced a stronger ion intensity on the positive polarity mode, while the Pt and Ir (detailed Fig. 3(a)–(c), 5(a)–(c)) ion counts are better observed on the negative polarity mode. Notably, CoB is of relatively low intensity in both polarities and thus B is neglected as an individual ion count in our analysis. In view of Pt and Ir layers, over Ta layer, being critical to the intermixing-induced alloy formation, we focused our analysis on negative polarity mode TOF-SIMS with stronger Pt and Ir profiles (detailed Fig. 3(a)–(c), 5(a)–(c)). TOF-SIMS depth profile analysis of Pt/CoB/Pt (Fig. S4(a)) shows approximately similar extent of intermixing at both Pt/CoB and CoB/Pt interfaces.

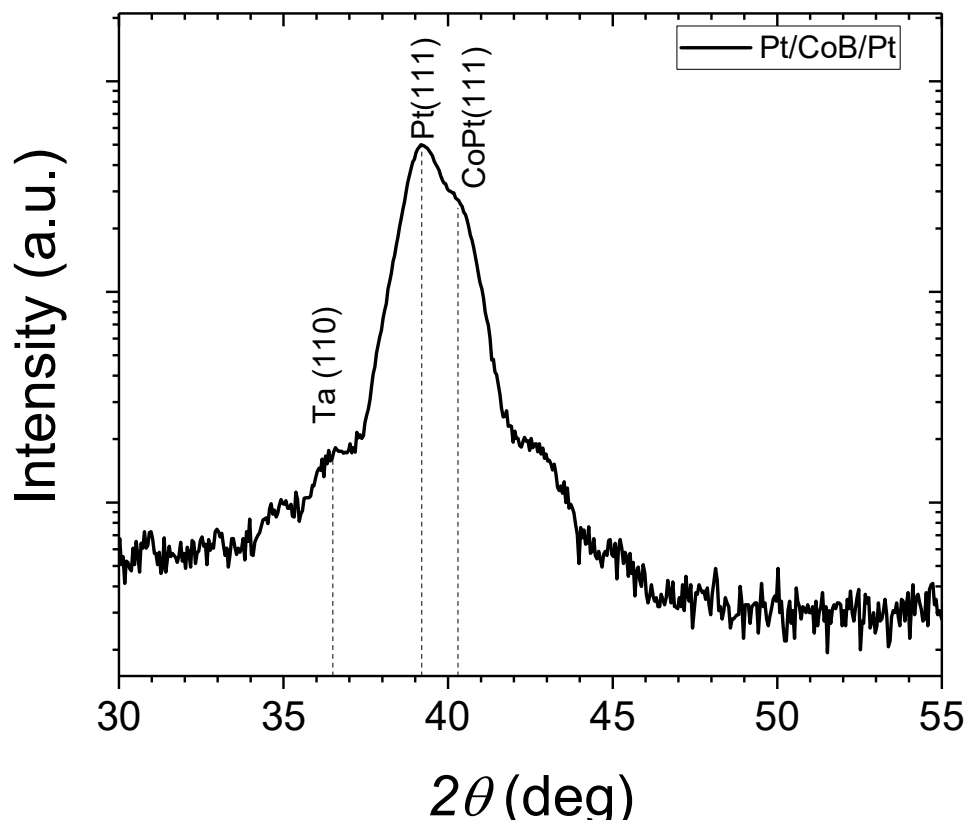

**Figure S5**. **XRD analysis for as-deposited Pt/Co/Pt multilayer stack**. X-ray diffraction (XRD) $\theta$-$2\theta$ spectrum for scan range of 30-55°, with step sizes of 0.05°, for the representative sample Pt/CoB(2.0)/Pt.

X-ray diffraction (XRD) was performed on the as-deposited Ta(3)/Pt(5)/Co(2)/Pt(5) multilayer stack. The Bragg peaks at 37.0° and 39.5° are associated with the Ta body-centred cubic (*bcc)* (110) and Pt face-centred cubic (*fcc)* (111) underlayers, respectively. Moreover, the presence of the Co-Pt *fcc* (111) peak at ~40.8° indicates the likely formation of $CoPt_3$ binary alloy due to intermixing at the Pt and CoB interfaces (Fig. S4(a)).

## S4. XRD Crystallographic Characterization

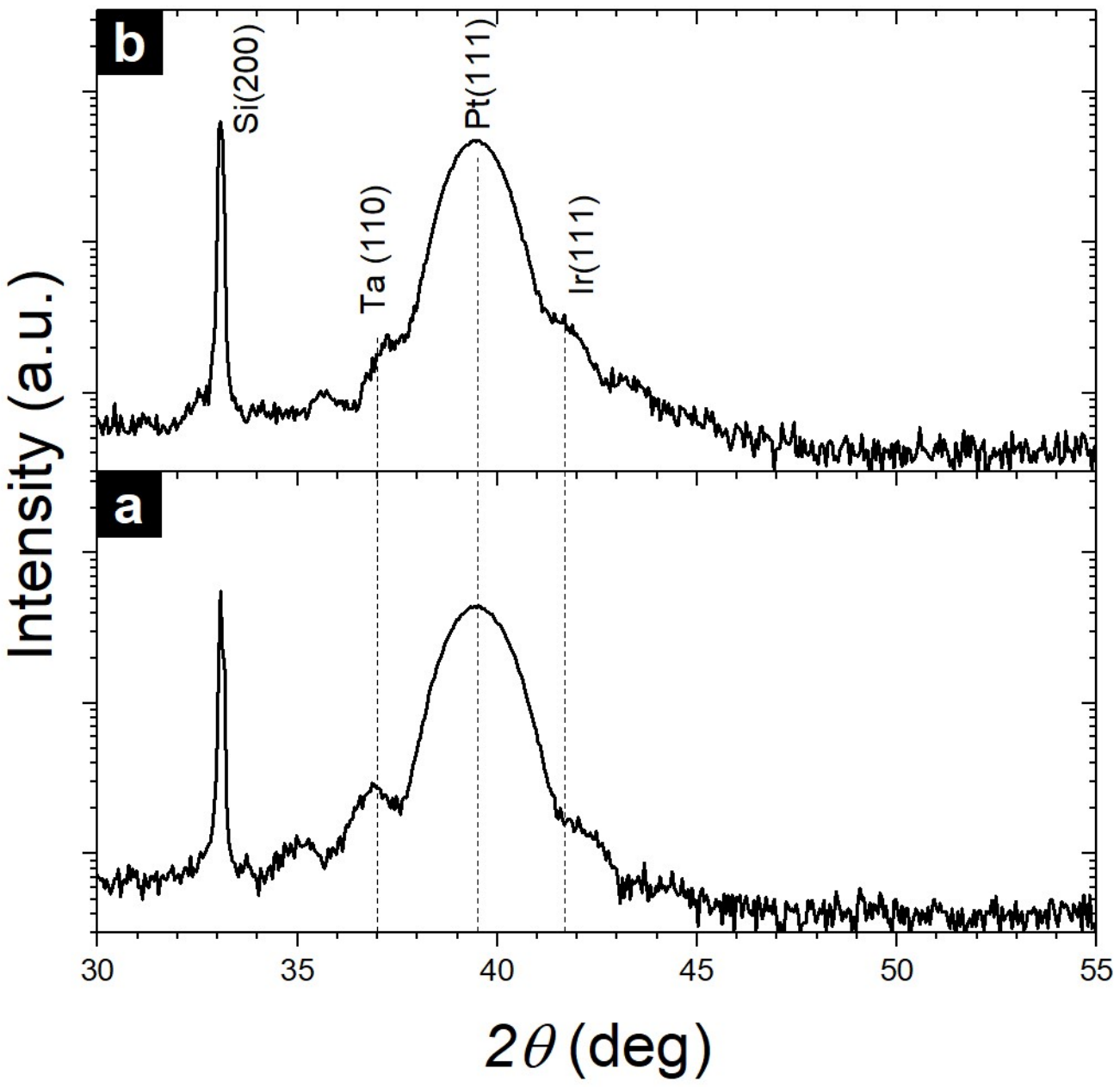

**Figure S6. XRD analysis of as-deposited Ta, Pt and Ir underlayers. (a-b)** X-ray diffraction (XRD) $\theta$-$2\theta$ spectra for scan range of 30–55°, with step sizes of 0.05°, for **(a)** Ta(3)/Pt(5), and **(b)** Ta(3)/Pt(5)/Ir(1) (thickness in nm in parentheses).

**XRD for Ta, Pt and Ir Underlayer.** Ta(3)/Pt(5) and Ta(3)/Pt(5)/Ir(1) (nominal layer thicknesses in nm in parentheses) were deposited on pre-cleaned thermally oxidized Si coupon. X-ray diffraction (XRD) was performed on these thin films to determine the crystallographic texture of Ta, Pt and Ir, providing clarity for identifying the remaining CoIr, CoPt and MgO peaks on the full multilayer stacks (Detailed in Fig. 3(d)–(f) and 5(d)–(f)). We identified *bcc* (110) and *fcc* (111) textures for the Ta and Pt underlayers at the Bragg peak positions of 37.0° and 39.5°, respectively. The Ir layer is too thin to be identified, whereby the expected Ir (111) peak appears overshadowed by the shoulder peak beside Pt (111).

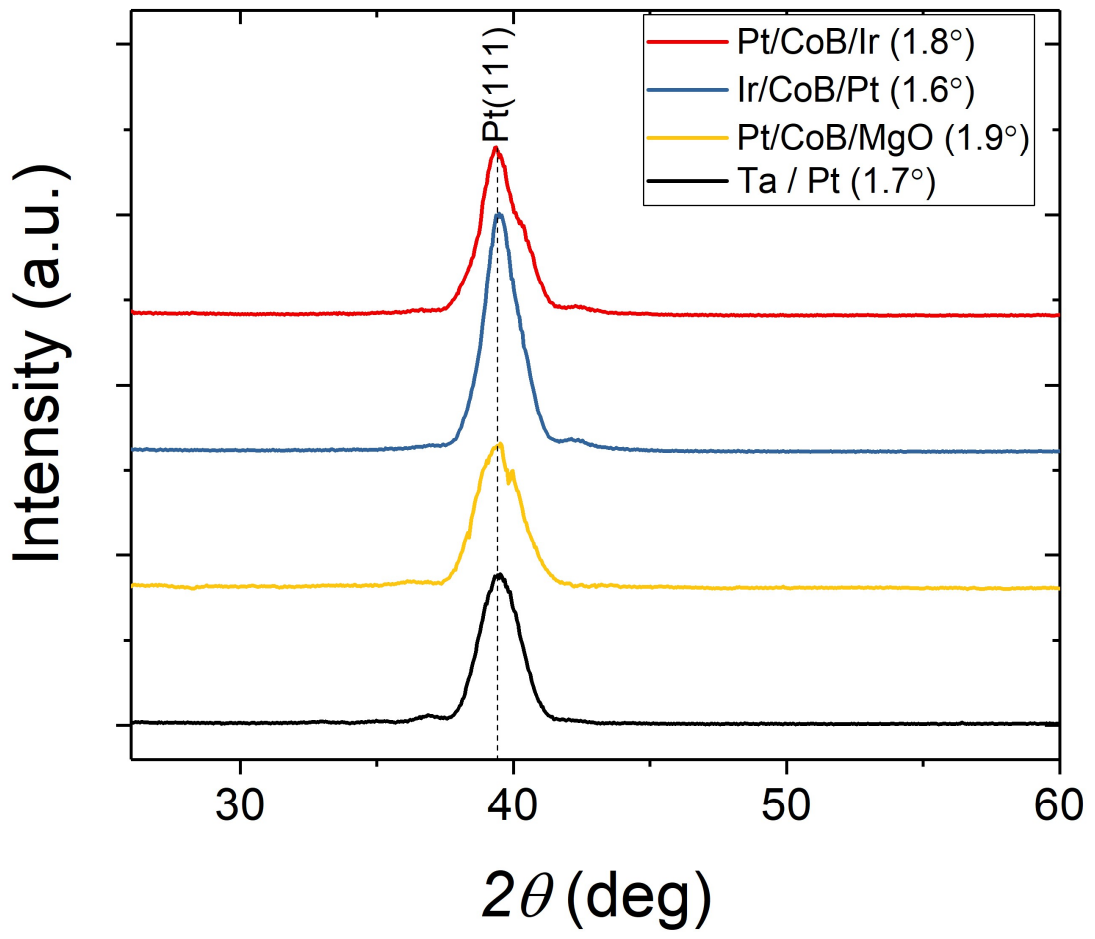

**Figure S7. XRD analysis for as-deposited reference layer and multilayer stacks.** XRD $\theta$-$2\theta$ spectra for scan range of 25 – 60°, with step sizes of 0.05°, for Ta(3)/Pt(5), Ta(3)/Pt(5)/CoB(2)/Ir(1)/Pt(4), Ta(3)/Pt(4)/Ir(1)/CoB(2)/Pt(4) and Ta(3)/Pt(4)/CoB(2.0)/MgO(1.5)/Pt(4) (thickness in nm in parentheses). The full width at half maximum (FWHM) of the Pt (111) Bragg peaks are indicated in the inset.

Fig. S7 shows the XRD $\theta$-$2\theta$ spectra, presented in linear scale, for the reference sample Ta(3)/Pt(5) and three multilayer stacks. The full width at half maximum (FWHM) of the Pt (111) Bragg peak across the three multilayers (1.6–1.9°) are constant to ~±10% with respect to the reference sample (1.7°), indicating borderline variations in the polycrystalline peak broadening/narrowing of the stacks relative to the reference Ta/Pt. The slight Pt peak broadening (~2°) in some stacks may be attributed to the weak presence of CoPt (111) crystallites. Meanwhile, the FWHM values are consistent with published results of polycrystalline Pt (111) thin films (~1.5°) deposited at room temperature using the physical vapour deposition technique.[1]

## S5. Fast Fourier Transform Analysis

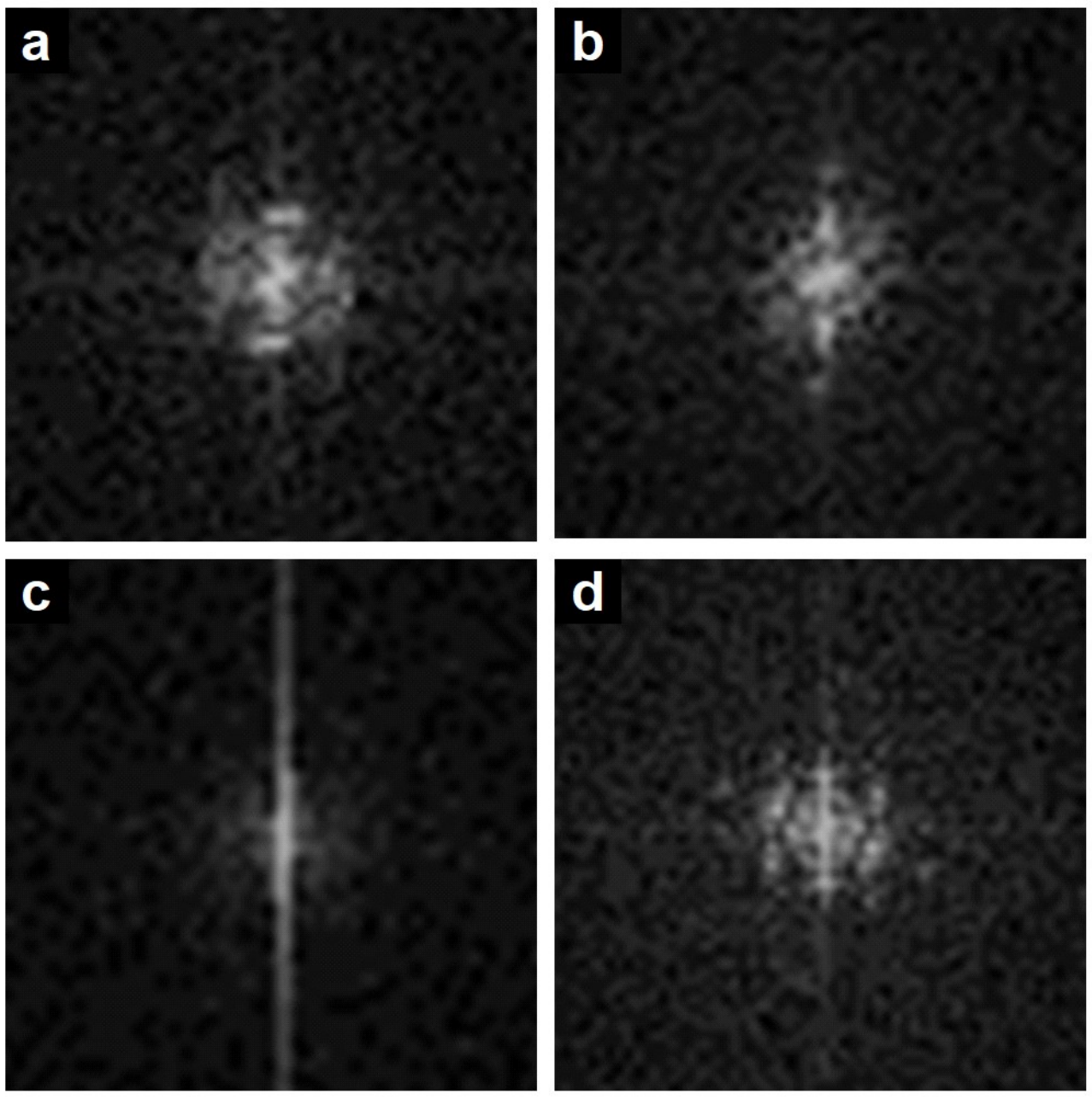

**Figure S8. HR-TEM analysis for as-deposited and annealed multilayer stacks.** Fast Fourier transforms (FFTs) of HR-TEM images of the CoB layer for as-deposited (a) Pt/CoB(2.8)/Ir, (b) Ir/CoB(2.0)/Pt (c) Pt/CoB(1.4)/MgO and (d) annealed Ir/CoB(2.0)/Pt multilayer stacks. For statistical averaging, (a)–(d) are obtained by summing 10 FFTs, each analysed over a ~2 × 2 nm region, across the CoB layer of each respective HRTEM image.

Cross-sectional high-resolution transmission electron microscopy (HRTEM) was performed using the Tecnai$^{TM}$ G2 F20 X-Twin. The TEM lamella samples were prepared with the standard focused ion beam lift-out technique using the FEI DA300 DualBeam$^{TM}$. To study the CoB crystallinity in the as-deposited and annealed multilayers, fast Fourier transform (FFT) patterns were obtained over 2 × 2 nm regions across the CoB layers in the HRTEM images using the Gatan Digital Micrograph$^{TM}$ software.

Across the as-deposited metallic multilayers (Fig. S8(a) & (b)), we observe bright diffuse spots in the CoB FFT patterns signifying the presence of polycrystalline CoB layer. XRD analysis (Fig. 3(d) & (e)) suggests that this polycrystalline phase corresponds to CoIr (002) and weak CoPt (111) crystallite formation. Upon annealing at 300 °C (Fig. S8(d)), the spots become more intense and less diffuse, indicating a more ordered crystalline phase in the CoB layer. Similarly, this is consistent with observations of increased annealing-induced alloying marked by stronger CoIr (002) and CoPt (111) XRD peak intensities (Fig. 5(d) & (e)). Meanwhile, the CoB FFT image for the Pt/CoB/MgO stack (Fig. S8(c)) shows weak spots, suggesting negligible CoPt (111) alloy content in a predominantly amorphous CoB layer.

## References

1. Karoutsos, V. *et al.* Growth modes of nanocrystalline Ni/Pt multilayers with deposition temperature. *J. Appl. Phys.* **102**, 043525 (2007).